\documentclass[pdftex,conference]{IEEEtran}

\usepackage{graphicx}
\usepackage{subfigure}
\usepackage{amsmath}
\usepackage{hyperref}
\usepackage{lipsum}
\usepackage{color}

\begin{document}
\title{Clustering network layers with the strata multilayer stochastic block model}

\author{
    \IEEEauthorblockN{Natalie Stanley\IEEEauthorrefmark{1}\IEEEauthorrefmark{2}, Saray Shai\IEEEauthorrefmark{2}, Dane Taylor\IEEEauthorrefmark{2}, Peter J. Mucha\IEEEauthorrefmark{2}}
    \IEEEauthorblockA{\IEEEauthorrefmark{1}Curriculum in Bioinformatics and Computational Biology} 
    %\\\ stanleyn@email.unc.edu}
    \IEEEauthorblockA{\IEEEauthorrefmark{2}
    Carolina Center for Interdisciplinary Applied Mathematics, Department of Mathematics, \\
    University of North Carolina, Chapel Hill
    %\IEEEauthorblockA{\IEEEauthorrefmark{3}
    %Statistical and Applied Mathematical Sciences Institute, \\
    %Research Triangle Park, North Carolina
    \\ stanleyn@email.unc.edu, \{sshai, taylordr\}@live.unc.edu, mucha@unc.edu}     
}

\maketitle

\begin{abstract}
Multilayer networks are a useful data structure for simultaneously capturing multiple types of relationships between a set of nodes. In such networks, each relational definition gives rise to a layer. While each layer provides its own set of information, community structure across layers can be collectively utilized to discover and quantify underlying relational patterns between nodes. To concisely extract information from a multilayer network, we propose to identify and combine sets of layers with meaningful similarities in community structure. In this paper, we describe the ``strata multilayer stochastic block model'' (sMLSBM), a probabilistic model for multilayer community structure. The central extension of the model is that there exist groups of layers, called ``strata'', which are defined such that all layers in a given stratum have community structure described by a common stochastic block model (SBM). That is, layers in a stratum exhibit similar node-to-community assignments and SBM probability parameters. Fitting the sMLSBM to a multilayer network provides a joint clustering that yields node-to-community and layer-to-stratum assignments, which cooperatively aid one another during inference. We describe an algorithm for separating layers into their appropriate strata and an inference technique for estimating the SBM parameters for each stratum. We demonstrate our method using synthetic networks and a multilayer network inferred from data collected in the Human Microbiome Project.  

%Given a multiplex network, it is of interest to aggregate layers that exhibit sufficient similarities in mesoscale structure. In this paper, we develop an approach to separate individual layers in to \emph{strata}, where we assume that members of a particular stratum have the same underlying generative model. In particular, the generative models are assumed to be stochastic block models. Using a variational expectation maximization approach, our method can effectively estimate the parameters for each stratum. We show that our method works on synthetic and real multilayer networks. Further, we elucidate the benefits of a mescale-based clustering approach to a fine-grained clustering approach in networks. 

%\boldmath
%Among layers in a multiplex network, there are inherent similarities, or correlations between layers. Prominent similarities can exist with respect to community structure. Under the assumption that groups of similar layers have the same underlying generative model, we present a method to simultaneously segregate layers in to homogeneous groups, or \emph{strata}, and to estimate the parameters of the underlying generative model. We show how using the mesoscale structure of networks is superior to relying on the fine-grain structure of the adjacency matrix in clustering tasks. Further, we provide results of numerical simulations to explore how clustering accuracy and model parameter inference evolve with varying community structure in synthetic data. Finally, we apply our method to real data sets in order to extract relevant information in real life applications. 
\end{abstract}

\begin{IEEEkeywords}
Stochastic Block Models, Clustering, Multilayer Networks, Strata, Probabilistic Models
\end{IEEEkeywords}

% \fi
\IEEEpeerreviewmaketitle

\section{Introduction}
Modeling relational information between a set of entities can often be successfully achieved through a network representation. Here, entities correspond to nodes and edges reflect some connection between them. In many applications, there are multiple ways to define an edge that can be collectively analyzed for a more thorough understanding of the data. Multilayer networks provide a framework to do this, in that each relational definition leads to a new layer in the network \cite{kivelamultilayer,boccaletti2014structure}.  Such data and corresponding networks have shown to be useful in many contexts, such as, in the comparison of genetic and protein-protein interactions in a cell \cite{genetic}, in understanding underlying relationships and community structure across social networks \cite{socialnetwork}, and in the analysis of temporal networks \cite{muchamultislice}. Thus, given the inherent mulitiplexity of network data across fields, there exists a need for the development of appropriate tools that can leverage information from all layers to elucidate structural patterns. \\
\indent Each layer in a multilayer network provides its own information about interactions between nodes, and it is useful to ask whether sets of layers are providing redundant information. Addressing this question requires the development of an approach to compress networks into a reduced-layer representation such that it effectively retains the information from the original multilayer network. Aggregating layers can potentially result in a loss of information, but it can also successfully corroborate the existence of underlying structural patterns. Moreover, this can lead to improved identification of structural patterns, including enhanced community detection \cite{airoldi}. This idea of reducibility in multilayer networks was previously explored in \cite{domen}: using an information{-}theoretic notion of distance between pairs of network layers, the authors performed hierarchical clustering of layers and chose the partition that maximized a quality function reflecting information loss due to the aggregation of layers. While this approach reflects the validity and usefulness of combining layers, it does not result in a generative model describing the clusters of redundant layers. To further this intuition to a probabilistic framework, we have developed the ``strata multilayer stochastic block model'' (sMLSBM), which seeks to compress the multilayer network by agglomerating sets of layers into structurally similar groups that we refer to as ``strata,'' while simultaneously clustering nodes into communities. To address this joint clustering question, our model assumes that network layers in a given stratum have the same underlying generative model for community structure. Importantly, because layers in a given stratum can be regarded as samples from a single probabilistic model, this can lead to improved community detection and parameter inference.

\subsection{Network Comparison Based on Community Structure}\label{sec:SBM}
\indent The problem of aggregating layers in a multilayer network is closely related to the problem of clustering networks. That is, given an ensemble of networks, one aims to identify sets such that networks within a set have similar characteristics. 
These characteristics, or ``features'' in this context, can describe any of the following: micro-scale structural properties such as subgraph motifs \cite{ugander2013subgraph,motiffinding}; multiscale properties such as community structure \cite{taxonomy}, the spectra of network-related matrices \cite{structurenetwork} and by defining latent roles \cite{netensemble}. Although clustering layers in a multilayer network is closely related to clustering networks in an ensemble, these are distinct problems with different difficulties and nuances. We focus on the prior pursuit; however, we expect for certain network ensembles that it will be beneficial to modify and apply our methods to the clustering of networks. 
%These include subgraph motifs \drt{\cite{scalablemotif,motiffinding}}, community structure \cite{taxonomy}, multiscale properties such as topological summaries [dane], and network summary statistics \cite{netensemble,structurenetwork}.
\\
\indent 
In this work, we analyze and compare layers in a multilayer network based on their community structure. Community detection in single-layer networks is an essential tool for understanding the organization and functional relatedness between nodes in a network \cite{porter2009communities,fortunato}.
Although there are many definitions for what constitutes a ``community'' \cite{rombach2014core}, one often assumes an ``assortative community'' in which there is a prevalence of edges between nodes in the same community as compared to the amount of edges connecting these nodes to the remaining network. In seeking to identify such communities, numerous approaches have been proposed, including those based on
%Identifying communities in networks typically requires the identification of the best partitioning of nodes into groups to maximize number of within-community edges, which can be quantified by multiple approaches, such as 
maximizing a modularity measure \cite{newmanmodularity} and fitting a generative probabilistic model \cite{abby}. Because each of these approaches present computational challenges for efficiently detecting communities, numerous heuristics exist for developing practical algorithms \cite{community,fortunato,leskoveccommunity,clausethierarchy,newmanspectral}.

%In this research we compare networks based on their community structure. Specifically, 
In seeking a statistically-grounded approach for studying communities in multilayer networks, we consider the stochastic block model (SBM) \cite{SBM}, a popular generative model for community structure in networks. The assumption of the SBM is that nodes in a particular community are related to nodes within and between communities in the same way, thus allowing SBMs to describe several types of communities (e.g., assortative, disassortative, core-periphery, etc. \cite{rombach2014core,aicher2015learning}). 
There are many other appealing aspects of stochastic block models; for example, a model-based approach allows for the denoising of networks through the removal of false edges and the addition of missing edges \cite{abby,guimera2009missing}.
The inference procedure for fitting SBMs to an undirected network with $N$ nodes and $K$ communities involves learning the two parameters, ${\boldsymbol \pi}$ and ${\bf Z}$. Parameter ${\boldsymbol \pi}$ is a $K \times K$ symmetric matrix, where $\pi_{mn}$ gives the probability of an edge existing between a given node in community $m$ and another node in community $n$. Matrix ${\bf Z}$ is an $N \times K$ indicator matrix, wherein each binary entry $Z_{im}$ indicates whether or not node $i$ is in community $m$. Each row of ${\bf Z}$ is constrained such that $\sum_{m=1}^{K} {Z}_{im}=1$, i.e. each node only belongs to 1 community. We also define vector $\boldsymbol z$, which has entries $z_{i}=\text{argmax}_m \{Z_{im}\}$ that indicate the community to which node $i$ belongs. For a given network, these parameters are often inferred through a maximum likelihood approach, and once learned, they provide information about the within and between community relatedness. 

\subsection{Related Work on Multilayer SBMs}

Due to the ubiquity of network data with multiple network layers, community detection in multilayer networks constitutes an important body of research. Important directions include generalizing the modularity measure \cite{muchamultislice} and studying dynamics \cite{manlio2} for this more general setting. 
%
%Of particular interest is recent research extending the SBM framework to multilayer networks
%
%\indent Providing an alternative to other methods for identifying communities in multilayer networks (e.g., maximizing multilayer modularity~\cite{muchamultislice}, or using multilayer modular flows \cite{manlio2}), there have been many recent developments in related multilayer stochastic block models (\cite{airoldi},\cite{catala},\cite{barbillon},\cite{thiagomlsbm},\cite{mlsbm1}). 
%
Given the usefulness of SBMs for the understanding of node organization in single-layer networks, it is important to extend SBMs to the multilayer framework, and indeed this direction of research is receiving growing attention  \cite{airoldi,mlsbm1,barbillon,catala,thiagomlsbm}. In this context, the general assumption is that there are shared patterns in community structure across the layers of a multilayer network, and the goal is to define and identify a stochastic block model that captures this structure. These works have explored many types of applications that can arise involving multilayer networks,
and have therefore given rise to several complementary models for multilayer stochastic block models (MLSBMs). We now briefly summarize this previous work that is very related, but notably different, from the model we study herein.
%
%Specifically, research thus far focused on two problems: one has a given network with many layers, or one has a single layer that is assumed to be derived from some aggregation of the layers. 
\\\indent 
 In Refs.~\cite{airoldi,mlsbm1,barbillon}, the authors studied situations in which many layers follow from a single SBM. In these instances, it is possible to obtain improved inference of the SBM parameters by incorporating multiple samples from a single model. For example, in Ref.~\cite{airoldi} the authors considered an increasing number of layers, $L$, and explored asymptotic properties of the estimated SBM parameters. Specifically, they fit an SBM to each individual layer in a way that utilizes the information from all layers, and they showed convergence of these estimators to their true values as $L\to\infty$. %As expected, as the number of layers increases, so does the quality of inference. 
For a network with $L$ layers and $K$ communities in each layer, their approach requires  an estimate of the community assignment matrix ${\bf Z}^l$ and probability matrix ${\boldsymbol \pi}^l$ for each layer $l$, the latter of which involves learning $K(K+1)L/2$ parameters.
% fitting their model to an $N$-node network with $K$ communities requires learning an $N \times K$ indicator matrix of community assignments across layers and a $K \times K$ matrix of block model probabilities, ${\boldsymbol \pi}^{{l}}$ for each individual layer, $l$. So, for a multilayer network with $L$ layers, and $K$ communities, there are $K(K+1)L/2$ total parameters to learn due to each ${\boldsymbol \pi}^{l}$, for $l \in \{1,2,\dots, L\}$. Particularly, 
To this end, the authors extended the variational approximation for approximating the maximum likelihood estimates of SBM parameters introduced in single-layer SBMs introduced in \cite{Dudin} to the multilayer setting.
\\\indent Ref.~\cite{airoldi} was followed up by Ref.~\cite{mlsbm1}, wherein the authors 
 %denote the model in \cite{airoldi} as MLSBM (multilayer stochastic model) and 
addressed issues that can arise for the model when $K$ and/or $L$ is large, or if the network is sparse.
 %problems with this approach as 
% the number of communities grows quickly or if layers are sparse overall. 
%To address these problems, t
They proposed a modified model called the restricted multilayer stochastic block model (rMLSBM). 
In this model, instead of learning a set of $L$ independent parameters, ${ \pi}^l_{mn}$, for each pair, $(m,n)$, each entry in ${\boldsymbol \pi}$ is fully layer-dependent so as to produce a reduction in the number of free parameters. Specifically, to determine the probability of an edge between a node from community $m$ and a node from community $n$ in layer $l$, they use a logistic link function and model the probability as $\text{logit}({ \pi}_{mn}^{l})={\pi}_{mn}+\beta_{l}$. The $\beta_{l}$ is an offset parameter representing the particular layer or type of edge. In this model, it is necessary %in an $L$ layer \drt{network} with \drt{$K$} communities
to learn $K(K+1)/2+L$ total parameters. Thus, the maximum likelihood estimate for an rMLSBM is a regularized estimator.\\
\indent  Consistent with the theme of fitting a single block model to a collection of layers, Ref.~\cite{barbillon} is similar to Refs.~\cite{airoldi} and \cite{mlsbm1} in that the authors seek to leverage information from all layers by considering the joint distribution of layers. Using this, they estimated quantities such as the marginal probabilities of node assignments to communities and the edge probabilities within and between groups. An interesting aspect of their approach is that they introduce a covariate capturing the coupling between pairs of nodes. For a network with $K$ communities and $L$ layers, this requires the estimation of $(2^{L}-1)K^{2}+(K-1)$ parameters. 
\\\indent We summarize Refs.~\cite{catala} and \cite{thiagomlsbm}, which provide techniques to determine whether a single layer network is the result of an aggregation procedure in a multilayer network. In Ref.~\cite{catala}, the authors defined a version of multilayer stochastic block model and an inference procedure for assessing whether or not a single-layer network was actually obtained from an aggregation of layers in a multilayer network; they considered the 
%they considered In particular, they \drt{seeked} to infer how network layer could have been 
aggregation of layers using boolean rules. Ref.~\cite{thiagomlsbm} describes two possible generative processes for multilayer networks: the \emph{edge-covariate} and \emph{independent-layer} models. In the edge-covariate model, an aggregated network is defined in which a given edge $(i,j)$ only appears in a single layer. Aggregating the layers in a multilayer network into a single network representation combines all of the edges from each of the layers. Thus, the translation of this idea into a generative model involves choosing a layer membership for each edge and sampling edges with a probability conditioned on adjacent nodes. In the independent-layer model, layers are generated independently from each other and the only constraint is that group membership of the nodes are the same across all layers.

\subsection{Contributions}
\indent 
While the literature on MLSBMs has recently grown quickly, there is still a need for a probabilistic generative model that allows for the layers in a multilayer network to be described by multiple SBMs. To this end, we
developed a novel multilayer stochastic block model, sMLSBM, that assigns network layers into disjoint sets that we call strata, where a collection of layers in a given stratum are assumed to be samples from the same underlying generative model. Our method can be viewed as a joint clustering procedure, where we seek to group layers into strata and nodes into communities. That is, we seek to simultaneously find layer-to-strata and node-to-community assignments. 
%
%Using node-to-community assignments \drt{across the all layers in a given stratum can allow one to more accurately estimate ${\bf Z}$ and ${\boldsymbol \pi}$ for that stratum,} and similarly a more accurate estimate of \drt{${\boldsymbol \pi}$} improves inference of ${\bf Z}$ for that stratum. 
%However, for S total strata and Ks possible communities that a node could belong to in a particular stratum s, there is a large combinatorial challenge associated with assigning nodes to communities and layers to strata. 
%To address this,
%
In order to address practical applications that can involve multilayer networks with several strata, layers, communities and nodes, we introduce an algorithm that effectively partitions layers into strata and an inference procedure to learn the SBM parameters for each stratum. Importantly, these two steps---assigning nodes to communities and layers to strata---are combined in an iterative algorithm so that an improvement in community detection can lead to an improvement in the clustering of layers into strata, which can iteratively lead to further improvement in community detection, and so on.
\\\indent To describe the model, the algorithm for fitting the model, and its performance, the remainder of this paper is organized as follows. In Sec.~\ref{sec:SMLSBM}, we define the model and an algorithm for fitting it. In Sec.~\ref{sec:Numerics}, we perform numerical experiments on synthetic networks. In Sec.~\ref{sec:Microbiome}, we test the model on correlation networks constructed from data from the Human Microbiome Project.

\begin{figure}
\begin{center}
\label{goals}
\includegraphics[width=1\linewidth]{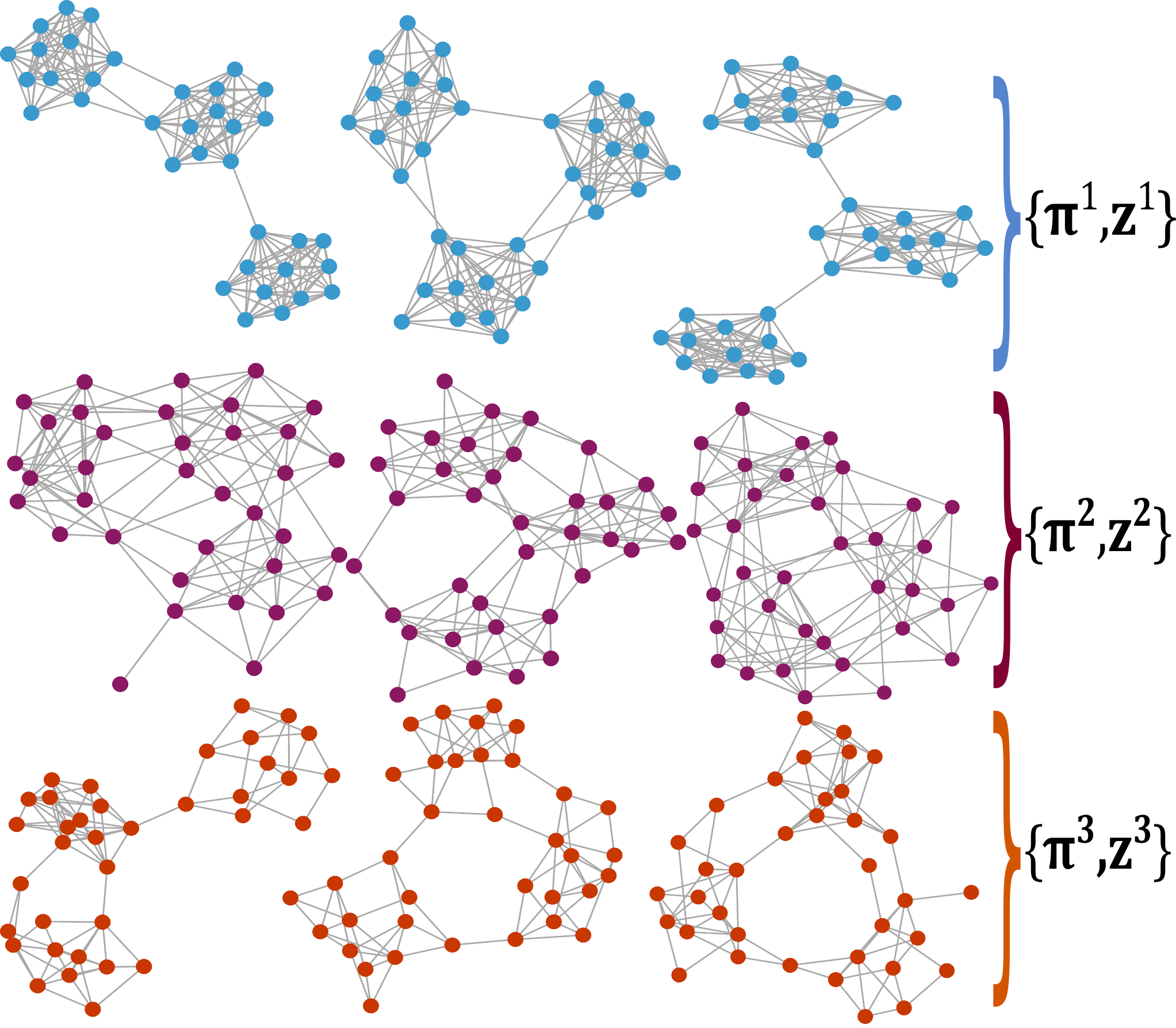}
\caption{{\bf Objective of strata multilayer stochastic block model (sMLSBM)}. Each of the $L=9$ networks here represents a layer in a multilayer network. Every network layer has $N=36$ nodes that are consistent across all layers. There are $S=3$ strata as indicated by the three rows and the colors of nodes. Clearly, network layers within a stratum exhibit strong similarities in community structure. That is, although each layer follows an SBM with $K=3$ communities, the SBM parameters are identical for layers within a strata but differ between layers in different strata. We would like to partition the layers into their appropriate strata and learn their associated SBM parameters, $\pi^s$ and $Z^s$.}
\end{center}
\end{figure}

\section{sMLSBM: Strata Multilayer Stochastic Block Model}\label{sec:SMLSBM}
\subsection{Network Definition}
Let $G(N,\mathcal{E})$ define a single network with N nodes and a set of undirected edges, $\mathcal{E}=\{(i,j)\}$. We define a multiplex network, which is one kind of multilayer network ~\cite{kivelamultilayer,boccaletti2014structure}, by defining a set of network layers, $G^{l}(N,\mathcal{E}^{l})$, where $l \in \mathcal{L}$ and the set $\mathcal{L}=\{1,2,\cdots,L\}$ indicates the layers' indices. We denote the collection of $L$ layers as a set, $\mathcal{G}$, such that $\mathcal{G}=\{{ {G}}^{1},{ {G}}^{2},\cdots,{ {G}}^{L}\}$ makes up the multiplex network and each element of the set is the network representing a layer. Furthermore, we define $\mathcal{A}=\{{\bf A}^{1},{\bf A}^{2},\cdots, {\bf A}^{L} \}$ to be the corresponding adjacency matrix representations of the network layers in $\mathcal{G}$. 
\subsection{Model Definition}
%For a network with $N$ nodes and $K$ communities, the objective in a traditional (single-layer) stochastic block model is to is to learn a $N \times K$ matrix, ${\boldsymbol{\pi}}$, and an $N\times K $ binary matrix $\bf{Z}$.
% Here, the parameters ${\boldsymbol{\pi}}$ and $\bf{Z}$ provide information about the distribution of edges within and between groups and the community memberships of each node, respectively. In particular, $\pi_{qt}$ represents the probability of an edge between a node in community $q$ and one in community $t$. $Z_{im}$ is an indicator variable for whether or not node $i$ belongs to community $m$ and $\sum_{m}{Z}_{im}=1$.  \\
 \indent Under the sMLSBM, the network layers, $G^{l}(N,\mathcal{E}^{l})$ are assumed to be generated by a set of $S$ stochastic block models, where the layers in stratum $s \in \{1,2,\cdots,S \}$, are parameterized by ${\boldsymbol \pi}^{s}$ and ${\bf Z}^{s}$ (or equivalently, vector ${\boldsymbol z}^s$, which has entries $z^s_{i}=\text{argmax}_m \{Z_{im}^s\}$ ). Note that the parameters ${\boldsymbol \pi}^{s}$ and ${\bf Z}^{s}$ for a single stratum are analogous in meaning to their respective parameters in the single-layer SBM case (see Sec.~\ref{sec:SBM}). 
For each stratum $s$, we let $\mathcal{L}^s\subseteq\mathcal{L}$ denote the set of layers corresponding to $s$, so that $\mathcal{L}=\bigcup_s \mathcal{L}^s$ and $\emptyset=\mathcal{L}^s\cap \mathcal{L}^t$ for all $s,t\in\{1,\dots,S\}$, $s\neq t$. We let $L^s=|\mathcal{L}^s|$ denote the number of layers in strata $s$ so that $\sum_s L^s=L$.
Finally, we allow the number of communities, $K^{s}$, to vary across the strata.
\\\indent
 %Furthermore, since a stratum is composed of multiple network layers, the parameters represent a consensus for that group.
For a given multilayer network, our objective during inference is to identify the stratum assignment of each layer and to learn the collection of strata parameters, $\boldsymbol{\Pi}=\{\boldsymbol{\pi}^{1},\boldsymbol{\pi}^{2},\dots,\boldsymbol{\pi}^{S}\}$ and $\mathcal{Z}=\{{\bf{Z}}^{1},{\bf{Z}}^{2},\dots {\bf{Z}}^{S}\}$. The learned SBM parameters for a stratum represent a consensus for the associated layers, and so in that sense can be interpreted as reducing the effective number of layers \cite{domen}. However, strata can also be interpreted as a way to simply identify layers with similarities in community structure. Figure 1 shows a toy example of a multilayer network with $S=3$ strata, where each layer has $N=36$ nodes and $K=3$ communities. Each individual network in this figure represents a layer in the network. The nodes in the layers belonging to each stratum are colored according to their stratum membership; moreover, it is easy to see that layers of a stratum exhibit high similarities in community structure.  \\
  \indent As part of our procedure, we specify another parameter that we refer to as the adjacency probability matrix, ${\boldsymbol \theta}^{s}$, which can be computed from $\boldsymbol{\pi}^{s}$ and ${\bf{Z}}^{s}$. Specifically, ${\boldsymbol \theta}^{s}$ is an $N \times N$ matrix such that ${\theta}_{ij}^{s}$ gives the probability of an edge between nodes $i$ and $j$ in stratum $s$. That is, ${\theta}_{ij}^{s}={ \pi}^{s}_{z_{i}^{s}z_{j}^{s}}$, where  $z_{i}^{s}$ specifies the community number for node $i$ in stratum $s$. Finally, we define the matrix ${\bf Y}$ of size $L\times S$, wherein an entry $Y_{ls}$ is a binary indicator of whether or not layer $l$ is assigned to stratum $s$. Note that $\sum_{s}Y_{ls}=1$. We also define a vector $\boldsymbol y$, which has entries $y_{l}=\text{argmax}_s \{Y_{ls}\}$ to indicate the strata to which layer $l$ belongs.

\subsection{Inference for sMLSBM}\label{sec:Algorithm}
\begin{figure}
\begin{center}
\includegraphics[width=1\linewidth]{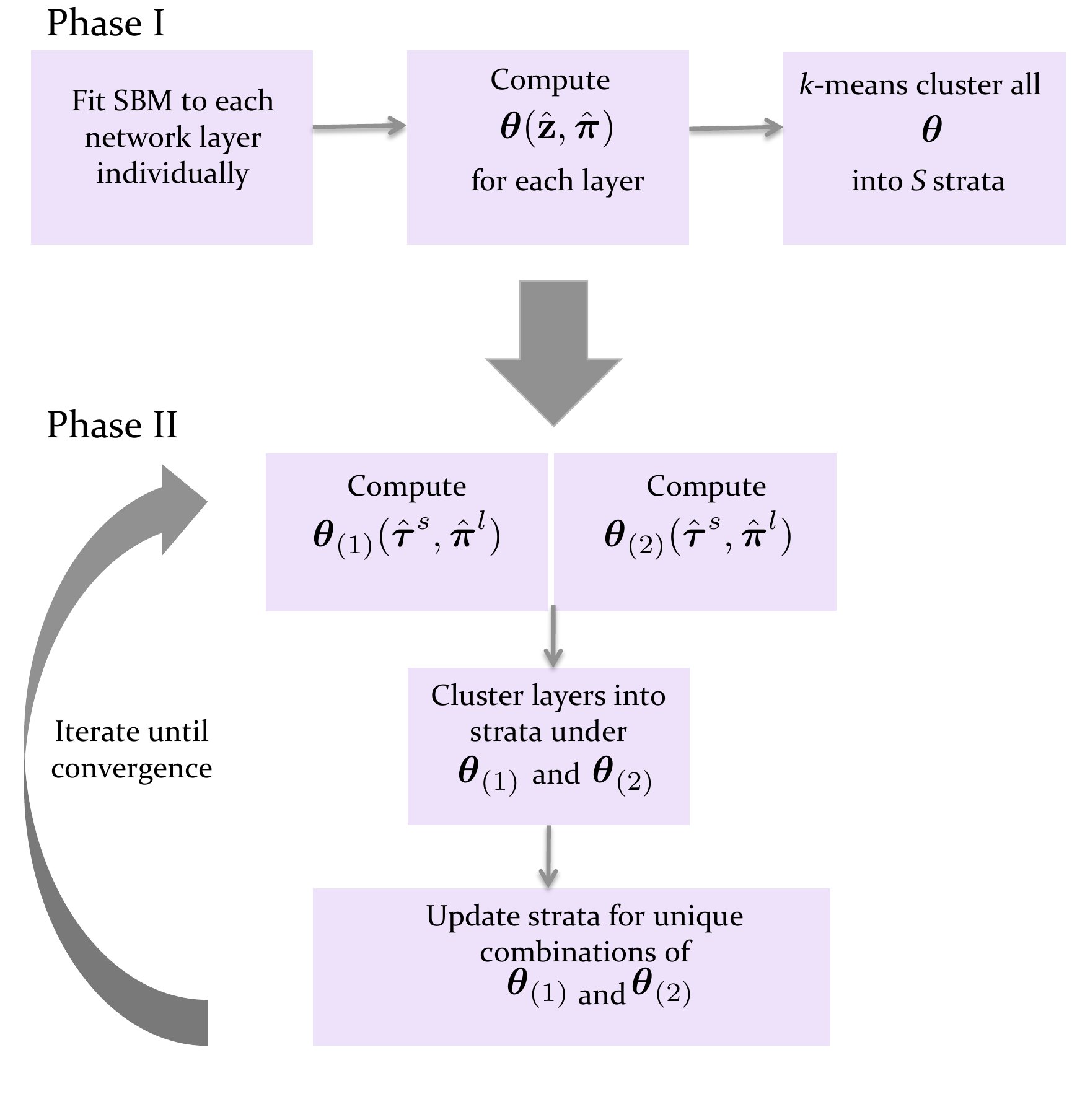}
\caption{{\bf Schematic illustration of our algorithm}: Our algorithm for fitting an sMLSBM is broken up into two phases: an initialization phase to cluster layers into strata, and an iterative phase that allows recursive learning between node-to-community and layer-to-strata assignments.}
\label{fig:Schematic}
\end{center}
\end{figure}

\indent The procedure for fitting an sMLSBM to a given network requires finding the layer-to-strata memberships and node-to-community memberships that best describe the multilayer network. For notational convenience, we introduce hat notation to represent the learned parameter estimate from the inference procedure. We can write down the marginal likelihood for the collection of network layers, $\mathcal{G}$, as,
\begin{equation}
\label{eq1}
p(\mathcal{G}\mid {\boldsymbol \Pi})=\sum_{\mathcal Z}\sum_{{\bf Y}}p(\mathcal{G},{\mathcal Z},{\bf Y} \mid {\boldsymbol \Pi}).
\end{equation}
We assume the probability of an edge between two nodes in layer $l$ belonging to stratum $s$ can be modeled as a Bernoulli random variable, based on the community membership of the nodes. In particular, $p({A}^{l}_{ij}=1)\sim \text{Bernoulli}({\bf \pi}_{z_{i}z_{j}}^{s})$. \\
\indent Since ${\bf Y}$ and ${\mathcal Z}$ are both latent quantities, searching over all possible values quickly becomes intractable. To tackle this issue, we develop a two-phase algorithm that incorporates a clustering algorithm for choosing the best $\bf Y$. This greedy approach leads to a significant reduction for the size of the search space since only $\mathcal Z$ must be statistically inferred. Specifically, during Phase I, we infer an SBM for each layer in isolation, and we cluster together sets of layers that have similar SBM parameters. Using these results as an initial condition in Phase II, we develop an iterative method that jointly identifies layer-to-stratum and node-to-community assignments as well as the SBM parameters for each stratum. We provide a schematic of the algorithm in Fig.~\ref{fig:Schematic}, and below we present the two-phase algorithm in detail.

\noindent{\bf Phase I.}
Phase I is comprised of two parts. First, we fit an SBM to each individual layer $l\in\{1,\dots,L\}$, which yields inferred SBM parameters $\hat{{\boldsymbol \pi}}^l$ and node-to-community memberships $\hat{{\bf Z}}^l$.
Then we cluster the layers based on the similarities of $\hat{{\boldsymbol \pi}}^l$ and $\hat{{\bf Z}}^l$. To infer $\hat{{\boldsymbol \pi}}^l$ and $\hat{{\bf Z}}^l$, we use the the inference method described in \cite{Dudin}. Here, the authors used a variational inference technique to approximate the maximum likelihood estimates for the stochastic block model parameters. For the set of $L$ layers, this produces sets of SBM parameters for each layer, which we denote by $\hat{\boldsymbol{\Pi}}=\{\hat{\boldsymbol{\pi}}^{1},\hat{\boldsymbol{\pi}}^{2}, \dots, \hat{\boldsymbol{\pi}}^{L}\}$ and $\hat{\mathcal{Z}}=\{\hat{\bf{Z}}^{1},\hat{\bf{Z}}^{2},\dots \hat{\bf{Z}}^{L}\}$ (that is, at this stage of the procedure, each layer is temporarily treated as its own stratum). Note also that each $\hat{{\boldsymbol Z}}^l$ can be equivalently represented by vector $\hat{{\boldsymbol z}}^l$, as described in Sec.~\ref{sec:SBM}. Using the estimates $\hat{{\boldsymbol \pi}}^{l}$ and $\hat{{\bf Z}}^{l}$ for a given layer, $l$, we can construct the corresponding adjacency probability matrix, $\hat{{\boldsymbol \theta}}^{l}$, which is defined entry-wise by $\hat{{\boldsymbol \theta}}_{ij}^{l}=\hat{\pi}_{\hat{z_{i}},\hat{z_{j}}}^l$. Doing this for each layer results in a collection of adjacency probability matrices, $\hat{\boldsymbol{\Theta}}=\{\hat{\boldsymbol{\theta}}^{1},\hat{\boldsymbol{\theta}}^{2}, \cdots, \hat{\boldsymbol{\theta}}^{L}\}$.\\
\indent Now, we seek an initial partition of layers into strata based on $\hat{\boldsymbol{\Theta}}$. The goal is to identify $S$ sets $\mathcal{L}^s$ so that the matrices $\{\hat{\boldsymbol{\theta}}^{l}\}$ with $l\in\mathcal{L}^s$ are close to one another, but they are distant from the remaining matrices, $\{\hat{\boldsymbol{\theta}}^{l}\}$ with $l\in\mathcal{L}\setminus \mathcal{L}^s$.
%
%where the total distance across strata between the stratum consensus adjacency probability matrix and the adjacency probability matrices of stratum member layers is as small as possible. 
This is accomplished by treating each $\hat{\boldsymbol{\theta}}^{l}$ as a feature vector and applying $k$-means clustering with $S$ centers so as to identify $S$ strata, $\mathcal{L}^s$. Note that $S$ can be selected \emph{a priori}, or approximated with a measure such as the gap statistic \cite{gap}. This gives us an initial estimate $\hat{{\bf Y}}$ for ${\bf Y}$. Note that this procedure initially treats each layer as a separate stratum, but provides a principled agglomeration of layers into $S\le L$ strata.   

\noindent{\bf Phase II.}
After a first-pass approach for assigning layers to strata, we initialize an iterative phase to more effectively estimate layer-to-strata assignments as well as the model parameters. Specifically, we would like to find the consensus SBM for each strata---that is, the $K^{s} \times K^{s}$ matrix $\boldsymbol{\pi}^{s}$ and the $N \times K^{s}$ matrix $\bf{Z}^{s}$ that maximize the likelihood of the observed layers in each stratum. We let ${\mathcal{A}}^{s}=\{{\bf A}^{l}\}$ for $l\in\mathcal{L}^s$ denote the collection of adjacency matrices corresponding to the $L^{s}$ layers in stratum $s$. 
%Hence each $\bf{A}$ is an $N \times N$ adjacency matrix. 

\indent We now proceed to maximize the likelihood in each stratum, by extending the framework of Ref.~\cite{Dudin} to a multilayer context. Note that this is similar to Ref.~\cite{airoldi}, except that we are not aiming to infer an SBM probability matrix for each layer, individually. In particular, the complete-data log-likelihood for stratum $s$ can be written as,
\begin{equation}
p({\mathcal{A}}^{s},{\bf{Z}}^{s})=p({\mathcal{A}}^{s}\mid {\bf{Z}}^{s})p({\bf{Z}}^{s}),
\end{equation}
where
\begin{equation}
p({\mathcal{A}}^{s} \mid {\bf{Z}}^{s})=\prod_{l \in \mathcal{L}^{s}}\prod_{i< j}\prod_{mn}{{{\pi}}^{s}_{mn}}^{A^{l}_{ij}}(1-{{\pi}}^{s}_{mn})^{(1-A_{ij}^{l})}.\
\end{equation}
To write $p({\bf{Z}}^{s})$, it is helpful to introduce a new parameter $\alpha^{s}_{m}$ that represents the probability that a randomly-selected node in stratum $s$ belongs to community $m$, i.e. $\alpha^{s}_{m} = p({{Z}}^{s}_{im}=1)$. 
%
%\todo{I DONT UNDERSTAND THIS!!! THE R.H.S. DEPENDS ON $i$ and should it be $p({{Z}}^{s}_{im}=1)$ ?Is it just the fraction of nodes in this community?? ??}
%
Note that $\sum_{m} \alpha^{s}_{m} =1$. 
%Further, we let $\mathcal{L}^{s}$ be the set of layers belonging to stratum $s$. 
Using this parameter, we can write
\begin{equation}
p({\bf{Z}}^{s})= \prod_{i}\prod_{m}\alpha{_{m}^{s}}^{({Z}^{s}_{im})}.
\end{equation}
It follows that the complete-data log-likelihood for the adjacency matrices representing the layers in stratum $s$ can be expressed as,
\begin{equation}
\begin{split}
\log P({\mathcal{A}}^{s},{\bf{Z}}^{s})&=\log(P({\bf Z}^{s}))+\log(P(\mathcal{A}^{s}\mid {\bf Z}^{s}))\\
&=\sum_{i}\sum_{m}{{Z}}^{s}_{im}\log(\alpha^{s}_{m})\\
&+\sum_{l \in \mathcal{L}^{s}}\sum_{i< j}\sum_{mn} A^{l}_{ij}\log({{\pi}}^{s}_{mn}) \\
&+\sum_{l \in \mathcal{L}^{s}}\sum_{i< j}\sum_{mn}(1-A^{l}_{ij})\log(1-{\pi}^{s}_{mn}).\
\end{split}
\end{equation}

Problems of this variety that involve the need to compute maximum likelihood estimates with incomplete data are typically addressed with the expectation maximization (EM) framework \cite{dempster}. Doing so requires the ability to compute $P({\bf{Z}}^{s}\mid{\mathcal{A}}^{s})$;  however, Ref.~\cite{Dudin} showed that it is intractable to calculate the conditional distribution for the single-layer network case. To address this challenge, we use a variational approximation, analogous to approaches in  \cite{airoldi,barbillon,Dudin}. In general, a variational approximation seeks to optimize a lower bound on the log-likelihood. To do this, we first approximate the conditional distribution, $P({\bf{Z}}^{s}\mid{\mathcal{A}}^{s})\approx{R}_{\mathcal{A}^{s}}$, where
\begin{equation}
R_{\mathcal{A}^{s}}({\mathbf Z}^{s})=\prod_{i}h({\mathbf Z}_{i\cdot}^{s};{\boldsymbol \tau}_{i\cdot}).
\end{equation}
Here, matrix ${\boldsymbol \tau}^{s}$ contains entries $\tau^{s}_{im}$ that approximate the probability that node $i$ belongs to community $m$ in stratum $s$. Further, function $h(\cdot)$ represents the multinomial distribution, with parameters, $\{{\boldsymbol \tau}^{s}_{im}\}$ for $m\in\{1,\dots,K^s\}$. Using this, we define the variational approximation as
\begin{equation}
\mathcal{J}(R_{\mathcal{A}^{s}})=\ell \ell(\mathcal{A}^{s})-\text{KL}(R_{\mathcal{A}^{s}}({\bf Z}^{s}),P({\bf Z}^{s}\mid \mathcal{A}^{s})),
\end{equation}
where $\ell \ell$ is log likelihood and KL is the Kullback-Leibler divergence. 

Through maximizing $\mathcal{J}(R_{\mathcal{A}^{s}})$, we minimize the KL divergence between the true conditional distribution, $P({\bf Z}^{s}\mid \mathcal{A}^{s})$, and its approximation, $R_{\mathcal{A}^{s}}({\bf Z}^{s})$. Moreover, we follow the derivation in Ref.~\cite{Dudin} and rewrite $\mathcal{J}(R_{\mathcal{A}^{s}})$ as
%When R_{\mathcal{A}^{s}}({\bf Z}^{s}) and P({\bf Z}_{s}\mid \mathcal{A}^{s}) are equal, the KL divergence is 0, and 
% In the variational approach, the goal is to optimize a lower bound of the observed data log-likelihood. In particular, we seek an approximation, $\mathcal{J}(R_{{\bf{X}}_{s}})$, to the conditional distribution,$P({\bf{Z}}^{s}\mid {\bf{G}}^{s})$ such that the KL-divergence between the true distribution and the approximation is as small as possible. Again from $\cite{Dudin}$, we introduce a variational parameter, ${\boldsymbol{\tau}}^{s}$ for each stratum as well as ${\alpha}^{s}$. Here, ${\boldsymbol{\tau}}^{s}$ is an $n \times k$ matrix, where $\tau^{s}_{ij}$ gives the probability that node $i$ belong to community $k$ in stratum $s$. Further, the $q_{th}$ entry in the k-length vector  ${\boldsymbol \alpha}$, $\alpha^{s}_{q}$ gives the probability that a node is in community $q$ in stratum $s$. 
%
%Plugging our modified joint distribution in to Daudin's variational approximation, we obtain,
%
\begin{equation}
\begin{split}
\mathcal{J}(R_{{\mathcal{A}^{s}}})&=\sum_{i}\sum_{m}\tau^{s}_{im}\log(\alpha^{s}_{m})\\
&+\sum_{l \in \mathcal{L}^{s}}\sum_{i<j}\sum_{mn}\tau^{s}_{im}\tau^{s}_{jn}[A^{l}_{ij}\log({ \pi}^{s}_{mn})]\\
&+\sum_{l \in \mathcal{L}^{s}}\sum_{i<j}\sum_{mn}\tau^{s}_{im}\tau^{s}_{jn}[(1-A^{l}_{ij})\log(1-{ \pi}^{s}_{mn})]\\
&-\sum_{i}\sum_{m}\tau^{s}_{im}\log(\tau^{s}_{im}).\
\end{split}
\end{equation}

We can now differentiate $\mathcal{J}(R_{{\mathcal{A}^{s}}})$ with respect to each parameter---while using Lagrange multipliers to enforce constraints (i.e. probabilities summing to 1)---to compute the updates. Doing so yields the following, where the hat notation symbolizes the current best estimate for the given parameter:
\begin{equation}
\hat{{{\alpha}}}^{s}_{m}=\sum_{i}\hat{\tau}^{s}_{im}/N \,,
\end{equation}
\begin{equation}
\hat{\pi}^{s}_{qt}=\frac{\sum_{l \in \mathcal{L}^{s}}\sum_{i<j}\hat{\tau}^{s}_{im}\hat{\tau}^{s}_{jn}A^{l}_{ij}}{\sum_{l \in \mathcal{L}^{s}}\sum_{i<j}\hat{\tau}^{s}_{im}\hat{\tau}^{s}_{jn}}\,,
\end{equation}
\begin{equation}
{\hat{\tau}}^{s}_{im} \propto  \hat{\alpha}^{s}_{m} \prod_{l \in \mathcal{L}^{s}}\prod_{i<j}\prod_{n}[{\hat{\pi}}_{mn}^{s}{^{A^{l}_{ij}}}(1-{\hat{\pi}}^{s}_{mn})^{1-A^{l}_{ij}}]^{\hat{\tau}^{s}_{jn}} \,.
\end{equation}
To find the best estimates for $\hat{{\boldsymbol{\tau}}}^{s}$ and $\hat{{\boldsymbol{\pi}}}^{s}$, we alternate between updating $\hat{{\boldsymbol{\tau}}}^{s}$ and $\hat{{\boldsymbol{\pi}}}^{s}$ until convergence. When convergence has occurred, we refer to the resulting estimates as the consensus $\overline{{\boldsymbol \tau}^{s}}$ and $\overline{{\boldsymbol \pi}^{s}}$ for stratum $s$. Similarly, $\overline{{\boldsymbol Z}^{s}}$ represents the consensus indicator matrix of node-to-community assignments computed from $\overline{{\boldsymbol \tau}^{s}}$. Note that we use the bar notation to reflect that the particular parameter estimate is for a stratum, rather than for an individual layer. 

Since $\overline{{\boldsymbol \tau}^{s}}$ and $\overline{{\boldsymbol \pi}^{s}}$ are computed in terms of each other, we can use one of the consensus parameters to compute the other parameter in individual layers. 
In particular, using the fixed node-to-community assignments from $\overline{{\boldsymbol \tau}^{s}}$, we compute the maximum-likelihood SBM parameters  for a particular layer $l$, which we denote with a tilde and hence, $\tilde{\boldsymbol{\pi}}^l$ and $\tilde{\boldsymbol{\tau}}^l$. Similarly, for fixed $\overline{{\boldsymbol \pi}^{s}}$, we compute the node-to-community assignments $\tilde{\boldsymbol{\tau}}^l$. Such estimates allow us to determine whether or not the stratum consensus estimates are accurate estimates for the SBMs of individual layers of the stratum. 
More importantly, as we shall now describe, these layer-specific estimates allow us to design an iterative algorithm that allows for alternating between learning the node-to-community and layer-to-stratum assignments.

To this end, we represent each layer by the adjacency probability matrix, which we compute two different ways: letting ${\boldsymbol{\theta}}({\boldsymbol{\tau}},{\boldsymbol{\pi}})$ represent the adjacency probability matrix specified by ${\boldsymbol{\tau}}$ and ${\boldsymbol{\pi}}$, % being used to compute the adjacency probability matrix for layer $l$. Specifically, 
we define
\begin{equation}
{\boldsymbol{\theta}}^{l}_{(1)}={\boldsymbol{\theta}}^{l}(\overline{{\boldsymbol{\tau}^{s}}},\tilde{\boldsymbol{\pi}}^{l}) ,
\end{equation}
%
%with the ${\boldsymbol \pi}$ that provides the best match to layer $l$ using information about node-to-community assignments given by \drt{$\overline{{\boldsymbol \tau}^{s}}$.}
%
\begin{equation}
{\boldsymbol{\theta}}^{l}_{(2)}={\boldsymbol{\theta}}^{l}(\tilde{{\boldsymbol{\tau}}}^{l},\overline{{\boldsymbol{\pi}^{s}}})
\end{equation}
%
%with the ${\boldsymbol \tau}$ that provides the best match to layer $l$ using information about the stochastic block model probabilities given by ${\boldsymbol \pi}^{s}$.
Note that the first definition uses the strata-consensus estimate for ${\boldsymbol \tau}^s$ and a layer-specific estimate for ${\boldsymbol \pi}^s$, whereas the latter uses a layer-specific estimate for ${\boldsymbol \tau}^s$ and the strata-consensus estimate for ${\boldsymbol \pi}^s$.
%
%That is, for each layer in stratum $s$, $\boldsymbol{\theta}^{l}_{(1)}$ uses the consensus ${\boldsymbol \tau}$ computed for stratum $s$ and the ${\boldsymbol \pi}$ computed to best fit a particular layer. Conversely, ${\boldsymbol{\theta}}^{l}_{(2)}$ uses the consensus ${\boldsymbol \pi}$ from the stratum paired with the single-layer estimates for ${\boldsymbol \tau}$ to compute the adjacency probability matrix for each layer in the stratum. 

%\\\indent 
During Phase I, we identified strata by clustering the adjacency probability matrices for the $L$ layers using the $k$-means algorithm. We employ a similar procedure here, but instead of clustering $L$ matrices, we now cluster $2L$ matrices, since each layer is represented in two different ways. Moreover, clustering these $2L$ matrices yields two cluster assignments for each layer. Typically, both representations of a particular layer will receive identical cluster {assignments---that is, for a given $l$, ${\boldsymbol{\theta}}^{l}_{(1)}$ and ${\boldsymbol{\theta}}^{l}_{(2)}$ are assigned to the same cluster, or strata. However, an interesting case arises when the two representations induce different stratum assignments for a given layer, because this implies that there is disagreement between ${\boldsymbol{\theta}}^{l}_{(1)}$ and ${\boldsymbol{\theta}}^{l}_{(2)}$, which implies uncertainty in the strata assignment of that particular layer $l$.
% ${\boldsymbol \pi}$ and single-layer ${\boldsymbol \tau}$ (and vice versa) do not have sufficient agreement. 
Because our iterative algorithm requires each layer to be assigned to a single stratum (i.e., we do not allow for mixed membership of layers into strata), layers with mixed membership according to ${\boldsymbol{\theta}}^{l}_{(1)}$ and ${\boldsymbol{\theta}}^{l}_{(2)}$ must be dealt with in some way. To account for these situations, we define additional strata for each combination of membership that arises. For example, if there are several layers $\{l\}$ that are clustered into stratum 1 according to ${\boldsymbol{\theta}}^{l}_{(1)}$ and stratum 2 according to ${\boldsymbol{\theta}}^{l}_{(2)}$, then we define a new stratum that contains only these layers. We note that there exists a variety of options for handling layers with such mixed membership after applying $k$-means clustering to ${\boldsymbol{\theta}}^{l}_{(1)}$ and ${\boldsymbol{\theta}}^{l}_{(2)}$ (e.g., one could assign such a layer to a stratum at random); however, we leave open for future work the exploration of these other options.
%the total number of partition combinations induced by the two representations of each layer determines the number of strata in the next iteration.
%\\\indent

After a single pass of Phase II, which requires layer-to-strata assignments (which can be encoded by vector $\boldsymbol y$) as input, the algorithm yields (ideally) improved layer-to-strata assignments (as well as consensus estimates for the SBM parameters of the strata, $\overline{{\boldsymbol \tau}^{s}}$ and $\overline{{\boldsymbol \pi}^{s}}$). Therefore, Phase II involves iterating the above procedure until the layer-to-strata assignments do not change. We note that in principle, it is possible for new strata to arise in each iteration (i.e., because we create strata to avoid mixed membership of layers), and this can allow the number of strata to grow with each iteration; however, we did not observe this issue in any of our synthetic or real data experiments. As we will show in the following section, convergence is typically observed after just a few iterations (e.g., see, for example, the second row of Fig.~4). If such an issue arises, it may be helpful to bound the number of iterations in Phase II. 

%To further illustrate the process of sMLSBM,  \drt{Fig.~\ref{fig:Schematic}} provides a schematic illustration of the inference and clustering procedures used to fit the model. 

\section{Synthetic Data Experiements}\label{sec:Numerics}
\subsection{Comparison of sMLSBM to other SBM Approaches}\label{sec:SBM1}
To demonstrate a situation where the sMLSBM framework has a clear advantage over other models, we designed a synthetic experiment and compared the results to two different SBM approaches: i) fitting a single SBM to all of the layers (denoted ``single SBM''), and ii). fitting a stochastic block model to each layer individually (denoted ``single-layer SBM"). We generated a multilayer network, where each layer has $N=128$ nodes, $K=4$ communities and an expected mean degree of $c=20$ (i.e., every network layer is expected to contain $cN/2=1280$ undirected edges). We specified an sMLSBM with $S=3$ strata and 10 layers per strata, which resulted in $L=30$ total layers. We defined ${\boldsymbol \pi}^{s}$ for each stratum $s$ in terms of two parameters, $p_{in}^s$ and $p_{out}^s$, which give the within-community edge probabilities and between-community edge probabilities, respectively. That is, we define ${\boldsymbol \pi}^s_{mn}=p_{in}^s$ when $m=n$ and ${\boldsymbol \pi}^s_{mn}=p_{out}^s$ when $m\not=n$. It follows that the expected mean degree is given by $c=N(p_{in}^s + (K-1)p_{out}^s)/K$.
In our experiment, we select the following SBM parameters: $(p_{in}^1,p_{out}^1)=(0.6,0.0083)$; $(p_{in}^2,p_{out}^2)=(0.4,0.075)$; and $(p_{in}^3,p_{out}^3)=(0.125,0.167)$. 
%Because we keep $c$ fixed, this requires that $p_{out}^1=$, $p_{out}^2=$, and $p_{out}^3=$
%Given the mean degree for networks belonging to each stratum is  25, this gives corresponding values of 0.00625, 0.04375, and 0.06875 for $p_{out}$. 
In Fig.~3(A), we show an example network layer from each strata. Nodes are colored by their community assignments in stratum 1. Note that the node-to-community assignments are different in each stratum and that the extent of block structure decreases from stratum 1 to stratum 3.

\begin{figure}
\begin{center}
\label{fig:SynExp1}
\includegraphics[width=1\linewidth]{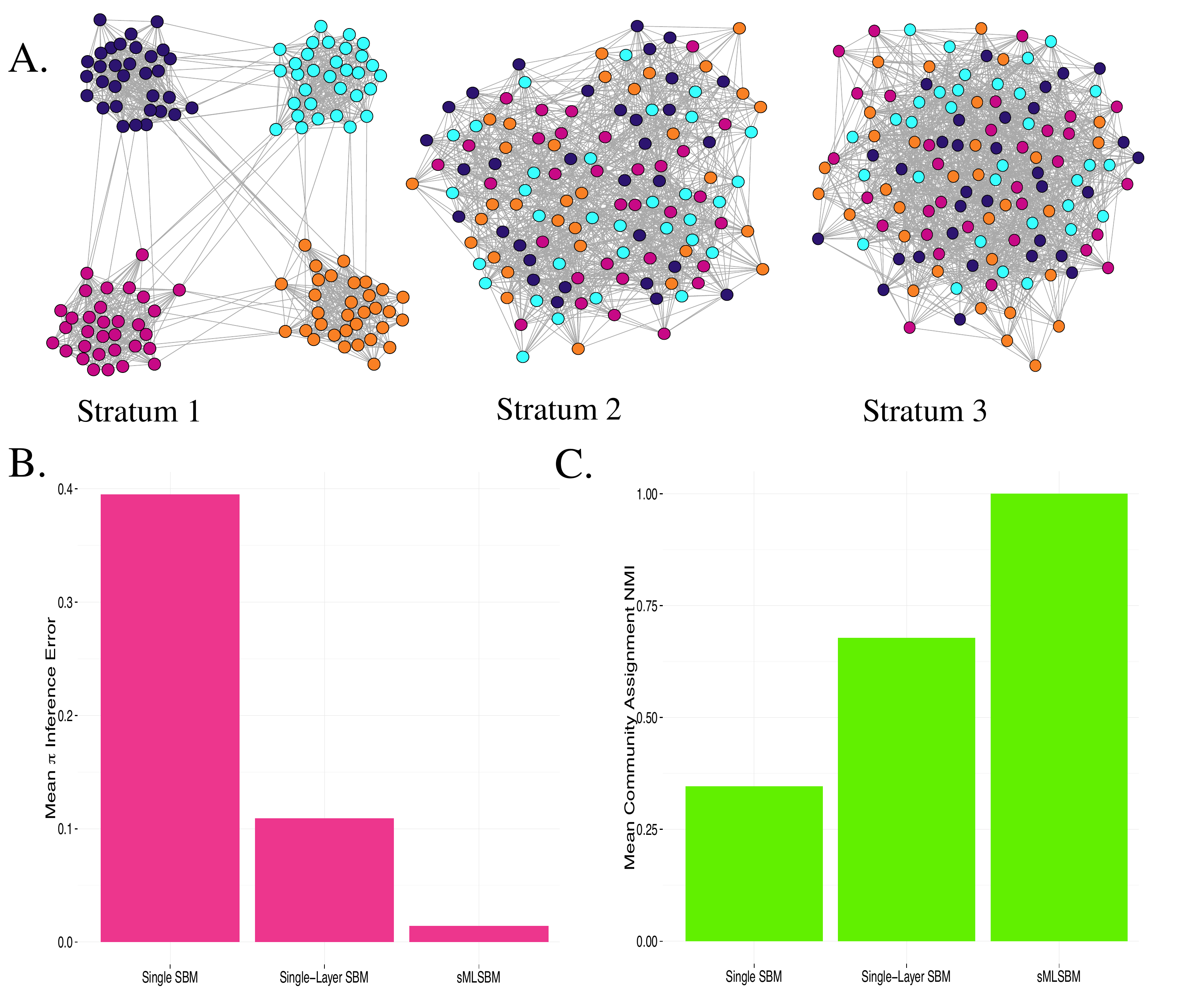}
\caption{{\bf \noindent Synthetic experiment comparing sMLSBM to other SBMs.} 
{\bf A}.~We specified a model with $S=3$ strata and $L=10$ layers per stratum. A representative layer from each stratum is plotted. Note that nodes in all networks are colored according to their community membership in stratum 1. Each network has $N=128$ nodes, $K=4$ communities and mean degree, $c=20$. The $p_{in}^s$ parameters for $s=1,$ $2$ and 3 are 0.6, 0.4 and 0.25, respectively. Corresponding values of $p_{out}^s$ were selected to maintain the desired expected mean degree, c=20. 
{\bf B}. We fit 3 types of models to the 30 network layers:
i) single SBM: fitting a single SBM to all of the layers;
ii) single-Layer SBM: fitting an individual SBM to each layer; and
iii) sMLSBM: identifying strata and fitting an SBMs for each strata. 
Each model yields an estimate $\overline{{\boldsymbol \pi}^{s_l}}$ for the true SBM of each layer $l$, which is denoted ${{\boldsymbol \pi}}^{l}$. Here $s_l$ denotes the inferred strata for layer $l$.
On the vertical axis we plot the mean $\ell$2 norm error 
%between each layer's true underlying ${\boldsymbol \pi}^{l}$ and that inferred given its stratum membership $s_{l}$ under the given model. In other words, we compute, 
$||\text{vec}({\boldsymbol \pi^{l})}-\text{vec}(\overline{{\boldsymbol \pi}^{s_{l}}})||_{2}$. 
 {\bf C}. For each of the three models, we computed the normalized mutual information (NMI) between the true node-to-community assignments ${{\bf z}^{l}}$ and the inferred values $\overline{{\bf z}^{s_l}}$.
 % inferred for \drt{every} layer under  \drt{every} layer under ${\bf Z}^{l}$ and its representation under the model, based on its indicator matrix for membership in stratum $s_{l}$, ${\bf Z}^{s_{l}}$. In other words, we compute $\text{NMI}(\drt{{\bf Z}^{s_l}}-{\bf Z}^{s_{l}})$.
   }
%as e%We fit 3 models to the 30 networks such that each model yields a representation, $\hat{{\boldsymbol \pi}^{l}}$ for layer $l$. The three models fit are 1) fitting an individual SBM to each layer, 2) fitting a single SBM to all of the layers, and 3) fitting SBMs based on strata memberships. On the vertical axis we plot the mean L2 norm error between each layer's true underlying ${\boldsymbol \pi}^{l}$ and that inferred under the given model. Strata SBM has the lowest mean error. }

%{\bf C}. }

%We also qualify the quality of node-to-community partitions within each layer under each model. For the community assignments inferred for each layer, $\hat{{\bf z}^{l}}$ under each of the 3 models, single layer SBM, single SBM and strata SBM, we computed the normalized mutual information (NMI) between the true ${\bf z}^{s}$ and $\hat{{\bf z}^{l}}$. Fitting an SBM individually to each layer and strata SBM greatly outperform the single SBM model.  
\end{center}
\end{figure} 

%\\\indent 
In order to compare the accuracy of fit for the three models---single-layer SBM, single SBM and sMLSBM---we quantify the inference accuracy of the SBM parameters, $\overline{{\boldsymbol \pi}^{y_{l}}}$, and community assignments, $\overline{{\bf Z}^{s_{l}}}$. 
First, for each layer and each model, we quantified the error ($\ell^{2}$ norm) between $\text{vec}(\overline{{\boldsymbol \pi}^{y_{l}}})$ and its true value, $\text{vec}({\boldsymbol \pi}^{l})$. Note that $\text{vec}({\bf X})$ is the $\frac{K(K+1)}{2}$ length vector representing the lower triangle of the matrix ${\bf X}$.  Moreover, to quantify error,
% parameter and the \drt{inferred parameter} $\overline{{\boldsymbol \pi}^{s_{l}}}$ under each of the \drt{three models.} 
%In other words, for each layer $l$ 
we compute $||\mbox{vec}({\boldsymbol \pi^{l}})-\text{vec}(\overline{{\boldsymbol \pi}^{s_{l}}})||_{2}$.  We note that this error is well-defined because we identify $K=4$ communities for all layers and all models. The mean error across layers under each model are shown in Fig.~3(B). In this example, sMLSBM outperforms the two other models.
Second, we computed for each layer the mean normalized mutual information (NMI) \cite{commdeccompare} between the true node-to-community assignments, ${\bf z}^{l}$, and the inferred values, $\overline{{\bf z}^{y_{l}}}$, under each model. In other words, for each layer, we compute, $\text{NMI}({\bf z}^{l},\overline{{\bf z}^{y_{l}}})$. Figure 3(C) shows the mean NMI for community assignments across layers.

\subsection{Synthetic Experiment with Two Strata}\label{sec:2strata}

Next, we further explored the performance of our algorithm (see Sec.~\ref{sec:Algorithm}) for inferring an sMLSBM under various situations: 1) in comparison to baseline clustering methods; 2) in response to an increase in the number of layers; and 3) under variations in levels of detectability. Specifically, we designed synthetic experiments in which we generated multilayer networks with either $L=10$ or $L=100$ layers. Every multilayer network contained $S=2$ strata (each having $K^1=K^2=4$ communities), and in each layer there were $N=128$ nodes (each having an expected mean degree of $c=16$). Note that in this example both strata have the same node-to-community assignments. The strata were fixed to be the same size, $L^1=L^2=L/2$. Similar to the experiment described in Sec.~\ref{sec:SBM1}, the SBM parameters were constructed using $p_{in}^s$ and $p_{out}^s$. Since we have already specified the expected mean degree, these parameters must satisfy the constraint $c=N(p_{in}^s+p_{out}^s)/2$ for both strata.
In all simulations, we fixed the SBM parameters of the first strata as $(p_{in}^1,p_{out}^1)=(.1836,.1055)$. It is also convenient to define the quantity, $N(p_{{in}}^{1}-p_{{out}}^{1})=10$, which relates to the detectability of communities \cite{decelle2011inference}. For example, the ability to detect community structure in a given layer and/or strata is, in general, expected to improve with increasing $N(p_{{in}}^{s}-p_{{out}}^{s})$. For the second strata, we allow $N(p_{{in}}^{2}-p_{{out}}^{2})$ to vary.

We present results for this experiment in Fig.~4, wherein the left and right columns give results for $L=10$ and $L=100$, respectively.
%In all panels, symbols indicate the mean values across 50 simulated multilayer networks, and error bars indicate the standard error.} 
%We show the results for 3 quantities \drt{of interest}: (A) accuracy of layer-to-strata assignments as measured through examining the NMI between ; (B) number of iterations required for sMLSBM to converge; and (C) accuracy node-to-community assignments.
% as a function of varying the quantity, $N(p_{{in}_{2}}-p_{{out}_{2}})$ for stratum 2. Thus, each experiment was defined by the particular value of $\text{Diff}_{2}$ using in 5 layers per stratum (left plot) and 50 layers per stratum (right plot). 
Symbols in each plot represent the mean over 50 multilayer networks, and error bars show standard error. In each plot, the vertical dotted line indicates $N(p_{{in}}^{2}-p_{{out}}^{2})=10$, which represents the point where the two strata are indistinguishable since $(p_{in}^1,p_{out}^1)=(p_{in}^2,p_{out}^2)$.
In Fig.~4(A), we show the NMI between the true layer-to-strata assignments and those inferred by sMLSBM, or $\text{NMI}({\bf y},\hat{\bf y})$. As a baseline, we compare sMSLBM results to directly clustering the layers' adjacency matrices using the $k$-means algorithm with $K=2$. We consistently observe higher NMI as a result of sMLSBM compared to $k$-means. More interestingly is the case with $L=100$, where both $k$-means and sMLSBM perform at least moderately well at partitioning layers into strata before the point where the strata are indistinguishable. %At this point, where the $p_{in}$ and $p_{out}$ parameters for strata 1 and 2 are the same, strata are clearly not distinguishable and we see this reflected in the drop of NMI. \\
In Fig.~4(B), we plot the number of iterations (NOI) required for Phase II of our algorithm to converge. We observe that as the number of layers in the network increases, so does the number of required sMLSBM iterations. Moreover, the peaks in panel B. correspond to the sudden jumps in strata NMI. 

%In both cases of $L=10$ and $L=100$, we notice a spike around when $N(p_{in}^{2}-p_{out}^{2})=20$.
%
Finally, in Fig.~4(C) we show the quality of node-to-community assignments 
%in \drt{the} strata. Particularly, we compute 
by plotting the NMI between the true and inferred node-to-community assignments as described in Sec.~\ref{sec:SBM1}. Note that stratum 1 here represents the stratum where the majority of layers were generated from model $S^{1}$ and analogously for stratum 2. Therefore, when the strata NMI is low (panel A.), we see poorer community detection results than expected, as layers get incorrectly mixed. As the strata NMI increases, layers from the same model are assigned together and the communities NMI stabilizes. 
%Specifically, we plot the mean NMI across stratum 1 (red symbols) and stratum 2 (blue symbols). As expected, we observe a general increase in NMI as $N(p_{{in}}^{2}-p_{{out}}^{2})$ increases. 
Finally, by comparing the results for $L=100$ to those for $L=10$, we observe an increase in number of layers, $L$, generally leads to an improvement in community detection and strata identification.\\

\begin{figure}
\begin{center}
\includegraphics[width=1\linewidth]{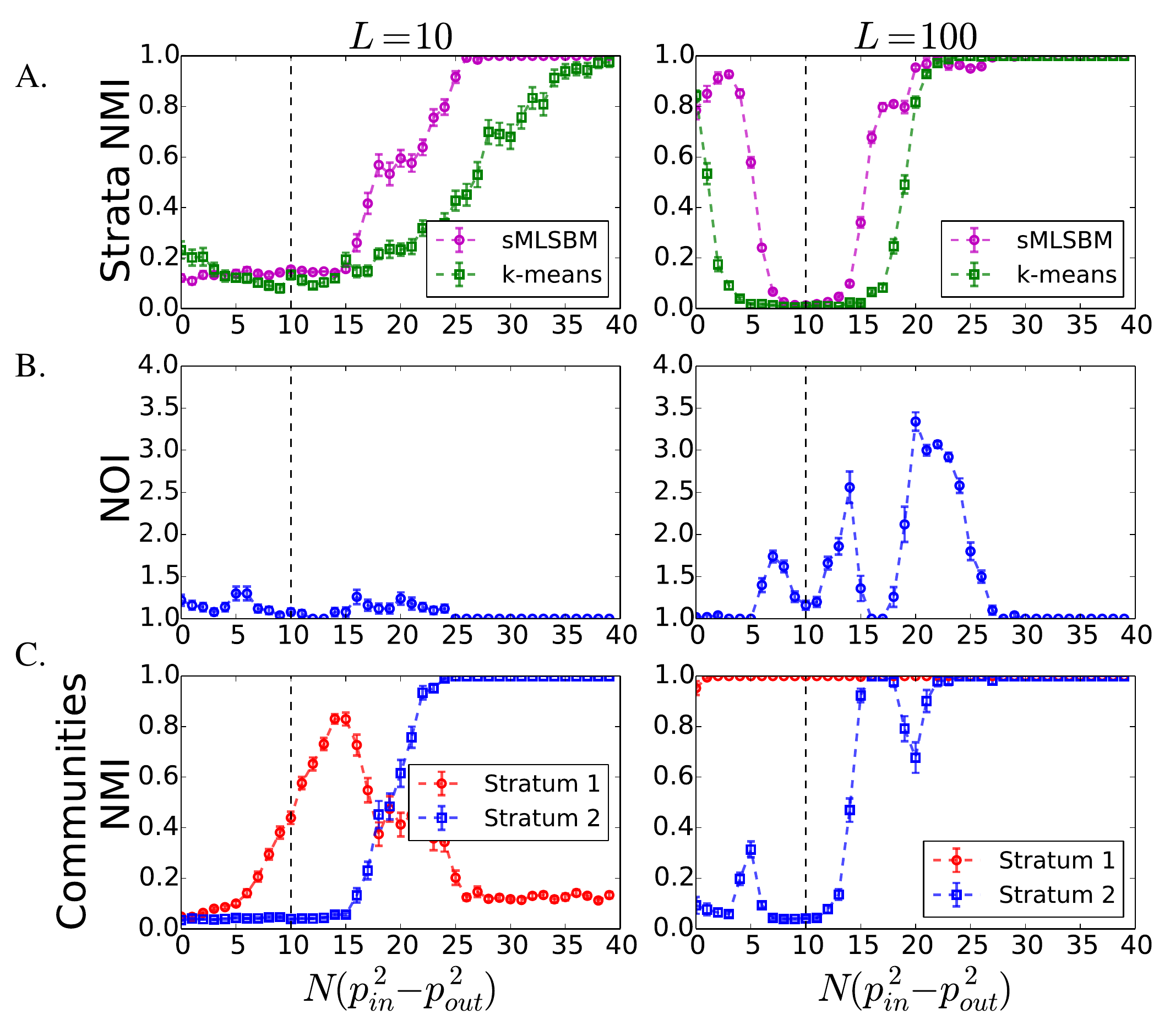}
\label{fig:saray}
\caption{
{\bf Synthetic experiment with two strata.} We conducted numerical experiments with multilayer networks with $N=128$ nodes, mean degree $c=16$, $S=2$ strata and $K^1=K^2=4$ communities. The networks contained either $L=10$ (left column) or $L=100$ layers (right column), which were divided equally into the two strata. For stratum 1, we fixed the quantity $N(p_{{in}}^{1}-p_{{out}}^{1})=10$, which fully specifies $(p_{{in}}^{1},p_{{out}}^{1})$ since setting $c=16$ also constrains these parameters. In contrast, we vary $N(p_{{in}}^{2}-p_{{out}}^{2})$.
%
%Then, in each experiment we simulate 50 2- stratum networks, where the $p_{in}$ and $p_{out}$ are fixed in stratum 1, but networks in stratum 2 are generated according to $\text{Diff}_{2}$ (horizontal axis). In each plot, error bars show standard error and curves are the mean from 50 simulated networks. The vertical line in the plots shows the point at which $\text{Diff}_{1}=\text{Diff}_{2}=10$. 
{\bf A}. As a function of $N(p_{{in}}^{2}-p_{{out}}^{2})$, we plot the mean NMI to interpret the ability of sMLSBM to recover the true layer-to-strata assignments. We compare the performance of sMLSBM (purple curve) to generic $k$-means clustering (green symbols) of adjacency matrices. 
{\bf B.} We plot the mean number of iterations (NOI) required for Phase II of our algorithm(see Sec.~\ref{sec:Algorithm}) to converge.
{\bf C.} Finally, we measure the quality of node-to-community assignment results by plotting the mean NMI between the true node-to-community assignments and those inferred with sMLSBM in stratum 1 (red symbols) and stratum 2 (blue symbols).}
%\caption{{\bf 2 Stratum Synthetic Experiment.} We considered numerical experiments consisting of multilayer networks with 200 nodes, 2 strata and 50 layers per stratum. Within-community edge probability, $p_{in}=.5$ for stratum 1 and a corresponding $p_{out}$ was chosen such that the mean degree, $c=50$. Numerical experiments consisted of  varying the within-community edge probability ($p_{in}$) for stratum 2, and measuring 3 quantities. Results shown correspond to mean and standard deviation obtained from 50 random networks. {\bf A.} As a baseline, we compared the performance of sMLSBM to $k$-means clustering of the adjacency matrices. Curves show the mean NMI across 50 simulations and error bars show standard deviation. {\bf B.} The mean number of iterations (NOI) required for sMLSBM to converge. {\bf C.} Each numerical experiment resulted in a node-to-community membership vector for each layer, depending on its strata assignment. We used NMI to compare this vector to the true node-to-community assignments in each layer. Here we have plotted the mean NMI across layers as a function of experimental parameter.}
\end{center}
\end{figure}
\section{Correlation Networks from the Human Microbiome Project}\label{sec:Microbiome}
\indent As an application of sMLSBM, we consider correlation networks constructed from data from the Human Microbiome Project \cite{microbiome}. For various sites on the body, the human microbiome project has successfully collected multiple human samples in order to better understand interactions between bacterial species. In this context, network inference is particularly interesting, as such methods aim to capture the relationships between various organisms. Microorganisms exhibit intricate ecologies within the gut of their human host and particular body sites have been shown to possess characteristic interactions. Further, certain interactions between microbes can often be associated with particular health and disease states \cite{microbeco}. Microbiome data is typically collected through metagenomic sequencing and reads are further binned into groups, known as operational taxonomic units (OTUs), to represent particular organisms. The nature of this count-based sequencing data makes network inference challenging, and is thus an interesting field in itself. To demonstrate the potential use for sMLSBM in the context of the human microbiome, we applied our algorithm for learning sMLSBMs to multilayer networks constructed from the SparCC \cite{sparcc} network inference method. \\
\indent SparCC is a correlation network inference method that aims to approximate the linear Pearson correlation between components in a system. This method performs favorably, as it accounts for the extent of diversity in the microbial community, which plays a significant role in detecting valid interactions. Furthermore, networks are constructed with the assumptions that the number of components in the system (e.g. OTUs) is large and that the correlation network should be sparse.  As supplemental data in Ref.~\cite{sparcc}, the authors provided their inferred microbial interaction networks for 18 sites in the human body. The edges in these networks have positive and negative real-valued weights, based on the results of SparCC inference. In this analysis, we converted the SparCC networks into binary adjacency matrices by allowing a link only if the SparCC edge-weight between two OTUs was at least 0.2 (as given in Ref \cite{sparcc}). To convert the 18 single-layer networks corresponding to species interactions in 18 body sites, we found the collection of nodes (OTUs) that occurred in at least 2 of the layers. This resulted in $N=213$ unique OTUs (nodes) for our multilayer network analysis.\\
\indent We inferred an sMLSBM for the multilayer network and found $S=6$ strata, implying that we find 6 clusters of body sites such that the microbiomes are similar between sites in the same cluster but differ from microbiomes at sites in the remaining clusters.
%\drt{As shown by the boxes in Fig.~5, this clustering} of layers (i.e., body sites) into strata \drt{did a very good job of identifying biologically relevant groups.}
%; for example, strata~1 contains only the left and right antecubital fossa, indicating that the microbiome of the `elbow pits' are similar to one another, but different from the microbiomes of other body sites.}
%was interesting because similar body sites tended to group together. 
To gauge the performance of our method, we compared the results to a hierarchical clustering (euclidean distance and complete linkage) of the networks. In Fig.~5  we show the dendrogram depicting the hierarchical clustering result, wherein the vertical axis denotes distance. Also captured in this figure are the sMLSBM results; leaves correspond to body sites and the colored boxes indicate strata assignments for sMLSBM, which do a very good job of identifying biologically meaningful groups.
%. We note that the orange, red, blue, green, and purple strata (as colored in the figure) are appropriate in terms of their location in the body. 
For example, it is intuitive that the saliva, hard palate and tongue dorsum layers have very similar microbe species interaction networks.  However, the stratum with $K^s=6$ layers seems to be a miscellaneous cluster. Using the dendrogram to compare the sMLSBM results with hierarchical clustering, we see that the quality of the clustering partition is highly dependent on where the tree is cut. It is difficult to find a cut of the tree that partitions the body sites in a way that is as meaningful as the result of fitting sMLSBM. Moreover, fitting sMLSBM also provides a generative model for each stratum. 
\begin{figure*}[t]
\begin{center}
\includegraphics[width=.6\linewidth]{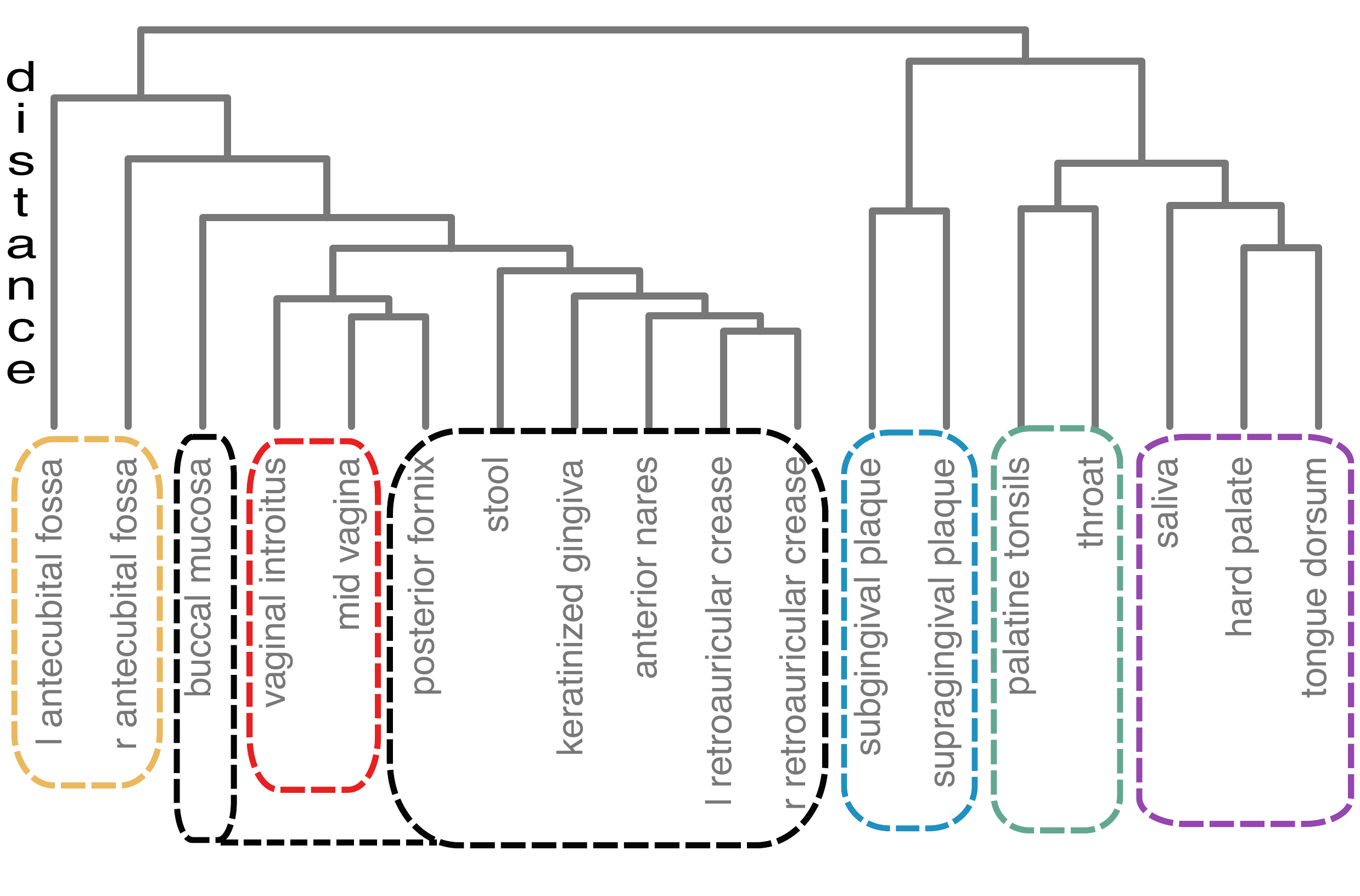}
\caption{
{\bf Hierarchical clustering of SparCC networks.} Hierarchical clustering was performed on binary adjacency matrices that were constructed by thresholding correlations between operational taxonomic units (OTUs), which give the nodes in our multilayer network. There are $L=18$ network layers that correspond to 18 body sites in the human body, as shown by the leaves in the dendogram. The data was obtained from Ref.~\cite{sparcc}. Colored boxes around the leaves indicate layer-to-assignments according to our fit of an sMLSBM. Although there is agreement between the hierarchical clustering and our inferred strata, we point out that there is considerable variability in the hierarchical clustering results as it is not possible to obtain a good partitioning of the layers into the sMLSBM-oriented strata by cutting the dendrogram horizontally. In contrast, there is a very clear biological relevancy of the strata inferred by sMLSBM (which moreover provides a generative model for the network layers).}
\end{center}
\end{figure*}
\\\indent The utility of having a probabilistic generative model for the microbiomes is illustrated in Fig.~6, where we illustrate network layers for 4 of the 6 strata that we identify. Specifically, each row provides information about the network layers and their fitted sMLSBM model for a particular stratum. Each grid in the figure represents the binary adjacency matrix encoding interactions between OTUs: a colored dot at position $(i,j)$ indicates the existence of an edge $(i,j)$ in the corresponding network layer.
%Edges, or a 1 in the adjacency matrix, are colored. 
In the first column of each row is a sample network generated with the learned SBM parameters of that stratum, $\overline{{\boldsymbol \pi}}^{s}$ and $\overline{{\bf Z}^{s}}$. Columns 2 and 3 show two representative network layers within the stratum. Note that while some strata have more than two members, for illustrative purposes we only show two example layers. 
It is easy to see the very similar block structure between all networks in a given row, corroborating the usefulness of the sMLSBM approach. 
\begin{figure}
\begin{center}
\includegraphics[width=1\linewidth]{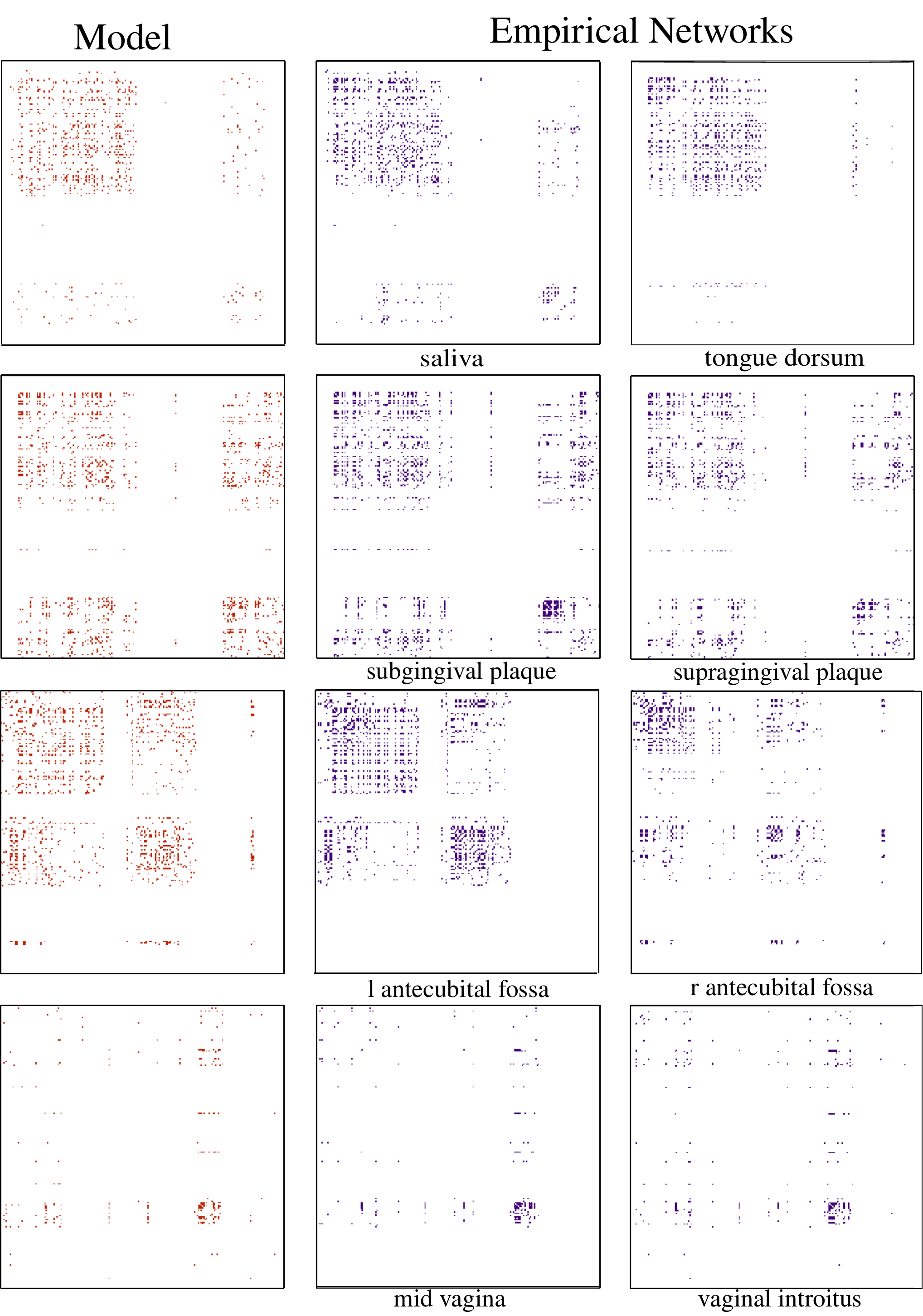}
\caption{{\bf Visualization of Strata in SparCC Networks.} We visualize the adjacency matrices for SparCC networks that encode microbiome interactions at body sites. In each panel, a colored dot at position $(i,j)$ indicates the existence of an edge $(i,j)$ in the corresponding network layer. The four rows correspond to four different strata. In column 1, we show a sample network generated from the SBM parameters, $\overline{{\boldsymbol \pi}^{s}}$ and $\overline{{\bf Z}^{s}}$, that we inferred for that stratum. In Columns 2 and 3, we show SparCC networks from that particular stratum. Note the strong similarity across each row.}
\end{center}
\end{figure}
%\begin{center}
%  \begin{tabular}{ l || c }
%	
%    {\bf Stratum} & {\bf Body Site Members} \\ \hline
%    1 & Anterior Nares,Buccal Mucosa, Keratinized Gingiva \\ \hline
%    %L Retroauricular Crease, Posterior Fornix, R Retroauricular Crease} 
%   2 & Hard Palate, Saliva,Tongue  \\ \hline
%   3 & L Antecubital Fossa, R Antecubital Fossa \\ \hline
%   4& Mid Vagina, Vaginal Introitus \\ \hline
%   5 & Palatine Tonsils,Throat \\ \hline
%    6 & Subgingival Plaque,Supragingival Plaque \\
%  \end{tabular}
%\end{center}

\section{Conclusion and Future Work}

\indent 
We developed a novel model for multilayer stochastic block models (MLSBMs) and an associated algorithm to jointly partition layers into strata and nodes into communities. Our model assumes that layers belonging to a stratum have community structure following the same underlying SBM. To fit sMLSBM to a multilayer network, and more-specifically, a multiplex network, we iteratively alternate between rearranging layer-to-strata assignments and updating the model parameters for each stratum. Having multiple networks within a stratum---hence multiple realizations from some underlying model---helps to make inference more accurate. Particularly, more accurate assignments of nodes-to-communities within a stratum leads to improved estimation of SBM probability parameters, and vice versa. 
We have shown for multiplex networks with several strata (e.g., see Fig.~3) that inaccuracies can arise if one attempts to fit a single SBM to the network or study the network layers in isolation.
%If layers from different models were all considered to have arisen from the same SBM, both the community memberships and SBM parameters used to represent each layer would be noisier and inaccurate. 
In contrast, our model allows for an understanding of the similarities between layers in a network, in terms of their community structure. The ability to identify strata within collections of networks holds promise in numerous applications. \\
\indent There are several extensions to sMLSBM that could make the approach more accurate and applicable to a wider range of applications. First, as is typical for SBMs, it would be useful to consider the degree-corrected \cite{degreecorrectSBM} and overlapping community (i.e., mixed-membership) \cite{degreecorrectSBM} varieties. Next, it may be useful to consider mixed membership of layers into strata, as well as nodes into communities. Further, sMLSBM as implemented here is only appropriate in unweighted, undirected networks. Extensions to weighted and directed networks, as shown in \cite{weightSBM} and \cite{sbmdirect}, could be quite useful. 
% we could consider the case where there exist layers that should not belong to any stratum and should be assigned to singleton clusters. 
\\
\indent Finally, the microbiome example considered here reveals some interesting computational biology questions that could facilitate the development of more advanced network tools. To construct the multilayer network, negative edges were thresholded away; however, antagonistic relationships between microbes are known to be important \cite{antagonism}. Thus, it would be useful to develop a signed version of sMLSBM that allows edges to be either positive or negative.
\\\indent The rise of a greater number of multilayer network datasets is providing the need for additional tools for the construction and analysis of such networks. The sMLSBM provides a new method to find signal in inherently noisy and complex network data.

\section*{Acknowledgments}
We thank James D. Wilson for helpful discussions about related work in multilayer networks, and in particular, multilayer stochastic block models. 
Research reported in this publication was supported by the Eunice Kennedy Shriver National Institute of Child Health \& Human Development of the National Institutes of Health under Award Number R01HD075712, the James S. McDonnell Foundation 21st Century Science Initiative Complex Systems Scholar Award grant \# 220020315, and training grants T32 GM 067553 and T32 CA 201159 from the National Institutes of Health. The content is solely the responsibility of the authors and does not necessarily represent the official views of the funding agencies.



%\bibliographystyle{ieeetr}
\bibliographystyle{IEEEtran}
%% argument is your BibTeX string definitions and bibliography database(s)
\bibliography{IEEbib2-drt}
%\begin{thebibliography}{1}
%\end{thebibliography}
%

\clearpage
\pagebreak

\end{document}